\documentclass[11pt, twocolumn, conference]{IEEEtran}
\usepackage[dvips]{graphicx}
\usepackage{multirow}
\usepackage{caption}
\usepackage{multicol,lipsum}
\usepackage{mathtools, cuted}
\usepackage{comment}
\usepackage{amsmath}
\usepackage{amsthm}
\usepackage[flushleft]{threeparttable}
\usepackage{array}
\usepackage{booktabs}
\newcommand{\PreserveBackslash}[1]{\let\temp=\\#1\let\\=\temp}
\newcolumntype{C}[1]{>{\PreserveBackslash\centering}p{#1}}
\newcolumntype{R}[1]{>{\PreserveBackslash\raggedleft}p{#1}}
\newcolumntype{L}[1]{>{\PreserveBackslash\raggedright}p{#1}}

\usepackage{bm}
\usepackage{graphicx,subfigure}
\usepackage{algorithmic, algorithm}
\usepackage{cite}
\usepackage{amsmath}
\usepackage{amssymb}
\usepackage{psfrag}
\usepackage{empheq}
\usepackage{latexsym, amsmath, subfigure, color, amsfonts, amssymb,graphicx}
\usepackage{wrapfig} 

\usepackage{tabularx}
\makeatletter
\def\hlinewd#1{%
\noalign{\ifnum0=`}\fi\hrule \@height #1 %
\futurelet\reserved@a\@xhline}
\makeatother
\begin{document}

\title{Deep Convolutional Compression for Massive MIMO CSI Feedback}

\author{Qianqian Yang, Mahdi Boloursaz Mashhadi and Deniz G\"{u}nd\"{u}z\\
\IEEEauthorblockA{Dept. of Electrical and Electronic Eng., Imperial College London, UK\\
Email: \{q.yang14, m.boloursaz-mashhadi, d.gunduz\}@imperial.ac.uk}}

\maketitle

\begin{abstract}
 Massive multiple-input multiple-output (MIMO) systems require downlink channel state information (CSI) at the base station (BS) to better utilize the available spatial diversity and multiplexing gains. However, in a frequency division duplex (FDD) massive MIMO system, the huge CSI feedback overhead becomes restrictive and degrades the overall spectral efficiency. In this paper, we propose a deep learning based channel state matrix compression scheme, called DeepCMC, composed of convolutional layers followed by quantization and entropy coding blocks. Simulation results demonstrate that DeepCMC significantly outperforms the state of the art compression schemes in terms of the reconstruction quality of the channel state matrix for the same compression rate, measured in bits per channel dimension. 

\end{abstract}
\section{Introduction}
Massive multiple-input multiple-output (MIMO) systems are considered as the main enabler of 5G and future wireless networks thanks to their ability to serve a large number of users simultaneously, achieving impressive levels of spectral efficiency. The base station (BS) in a massive MIMO setting relies on the downlink channel state information (CSI) to achieve the promised performance gains. Therefore, massive MIMO systems are more amenable to time division duplex (TDD) operation, which, thanks to channel reciprocity, does not require CSI feedback. Frequency division duplex (FDD) operation is more desirable due to better coverage it provides; however, channel reciprocity does not hold in FDD; and hence, downlink CSI must be estimated at user equipments (UEs) during the training period and fed back to the BS. 

The resulting feedback overhead becomes significant due to the massive number of antennas, and has motivated various CSI reduction techniques based on vector quantization \cite{CodebookCSI1, CodebookCSI2} and compressed sensing (CS) \cite{CSCSI1, CSCSI2}. In vector quantized CSI feedback, the overhead scales linearly with system dimensions, which becomes restrictive in many practical massive MIMO scenarios. On the other hand, CS-based approaches rely on sparsity of the CSI data in a certain transform domain, which may not represent the channel structure accurately for many practical MIMO scenarios. CS-based approaches are also iterative, which introduces additional delay.

More recently, following the recent resurgence of machine learning, and more specifically deep learning (DL) techniques for physical layer communications \cite{ MLintheAir}, deep learning (DL)-based CSI compression techniques have also received attention, and are shown to provide significant gains compared the aforementioned approaches in the literature  utilizing the sparsity prior.  
The CSINet scheme, proposed in \cite{wen2018deep}, is based on a neural network (NN) autoencoder architecture to obtain a compressed representation of the CSI by learning low-dimensional features of the channel gain matrix from training data. In \cite{DLCSI3}, the  authors improve CSINet by utilizing a recurrent neural network to utilize temporal correlations in time-varying channels as well. Utilizing bi-directional channel reciprocity, the authors in \cite{DLCSI4} use the uplink CSI as an additional input to further improve the results utilizing the correlation between downlink and uplink channels.   

In this paper, we develop a CSI compression network (DeepCMC) that  utilizes a deep  convolutional NN (CNN) in conjunction  with  quantization  and  entropy  coding  blocks  to  efficiently compress  and  encode  the  CSI  for  downlink  MIMO  channels. In comparison with the previous DL-based CSI compression techniques, the main contributions of the proposed DeepCMC architecture can be summarized as follows:

i) Existing DL-based architectures for CSI compression \cite{wen2018deep, DLCSI3, DLCSI4} all include a fully connected layer, which means that it can only be utilized for the specified input size, i.e., for a given number of transmit antennas and number of subchannels. This would mean that a different NN needs to be trained for each possible setting, and UEs need to store NN coefficients for all these networks, limiting the practical implementation of these solutions. On the other hand, the proposed DeepCMC architecture is fully convolutional, and has no densely connected layers, which makes it flexible for a wide range of MIMO scenarios with different number of sub-channels and antennas. As shown by the simulation results, although DeepCMC is trained for a specific number of sub-channels and antennas it shows negligible performance degradation for different number of sub-channels and antennas.
\begin{figure*}[h!]
\centering
\includegraphics[scale=0.5]{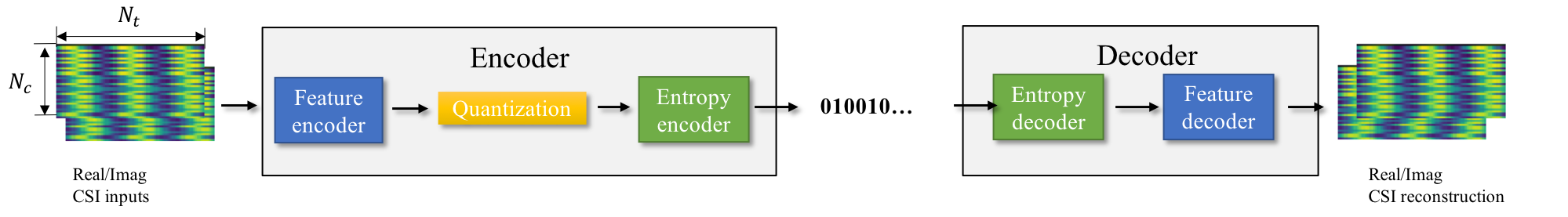}
\caption{The encoder/decoder architecture for the proposed CSI feedback compression scheme DeepCMC.}
\label{single}
\end{figure*}

ii) Previous works on NN-based CSI compression \cite{wen2018deep, DLCSI3, DLCSI4} mainly utilize autoencoder architectures for feature extraction, whose output is a low-dimensional complex vector. The elements of this vector are then sent to the transmitter by using a 32-bit representation for each component. Instead, DeepCMC includes the quantization and entropy coding blocks within its architecture to directly convert the channel gain matrix into bits for subsequent communication. In contrast to the literature that simply minimize the reconstruction mean square error (MSE), DeepCMC is trained with a rate-distortion cost that takes into account both the compression rate (in terms of bits per channel dimension) and the reconstruction MSE which significantly improves the achieved performance in comparison with the previous works.

We would also like to note that DeepCMC can be further improved utilizing similar approaches as in \cite{DLCSI3, DLCSI4} to benefit from temporal correlations or channel reciprocity.

This paper is organized as follows. In Section II, we present the system model. In Section III we present the proposed DeepCMC architecture for CSI compression. Section IV provides the simulation results and comparison with the state of the art results, and Section V concludes the paper.

\section{System Model}\label{sec1}
We consider a massive MIMO channel, where a BS with $N_t$ antennas serves a single-antenna user utilizing orthogonal frequency division multiplexing (OFDM) over $N_c$ subcarriers. We denote by $\mathbf{H}\in \mathbb{C}  ^{N_c \times N_t}$ the channel matrix, and by $\mathbf{v} \in \mathbb{C}^{N_t \times 1}$ the precoding vector. The received signal at the user is then given by
\small
\begin{align}
    \mathbf{y}= \mathbf{H} \mathbf{v} x + \mathbf{z},
\end{align}
\normalsize
where $x \in \mathbb{C}$ is the data-bearing symbol, and $\mathbf{z} \in \mathbb{C}^{N_c \times 1}$ is the additive noise vector.

In order to design the pre-coding vector $\mathbf{v}$ for efficient transmission, the BS uses an estimate of the CSI matrix values $\mathbf{H}$. To this end, in an FDD MIMO system, users estimate downlink CSI values through pilot-based techniques, and feedback the estimated CSI to the BS. However, the excessive overhead of CSI data, i.e., $N_c \times N_t$, particularly for massive MIMO systems, becomes prohibitive considering the limited feedback resources available in practical scenarios.

To cope with this challenge, efficient compression and encoding of the channel matrix $\mathbf{H}$ is desirable. Let $\mathbf{H} = [\mathbf{h}_1, \mathbf{h}_2, \hdots, \mathbf{h}_{N_c}]^T$, where $\mathbf{h}_n \in \mathbb{C}^N_t$ is the channel gain vector over subcarrier $n$, for $n= 1, \ldots, N_c$, and assume the BS is equipped with a uniform linear array (ULA) with response vector (simple 2D case):
\small
\[
\mathbf{a}(\phi)=[1, e^{-j\frac{2\pi d}{\lambda} \sin{\phi}}, \cdots, e^{-j\frac{2\pi d}{\lambda} (N_t -1) \sin{\phi}}]^T,
\]
\normalsize
where  $\phi$ is the angle of departure (AoD), and $d$ and $\lambda$ denote the distance between adjacent antennas and carrier wavelength, respectively. According to \cite{HMIMO}, the channel gain vectors can be modeled as
\small
\begin{align}\label{Hform}
    \mathbf{h}_n=\sqrt{\frac{N_t}{L}}\sum^{L}_{l=1} \alpha_l e^{-j2\pi \tau_l f_s \frac{n}{N_c}} \mathbf{a}(\phi),
\end{align}
\normalsize
where $L$ is the number of multipath components, $f_s$ is the sampling rate, $\tau_l$ is the delay, and $\alpha_l \sim \mathcal{CN} (0,\sigma^2_{\alpha})$ is the propagation gain of the $l^{th}$ path with $\sigma^2_{\alpha}$ denoting the average power gain. According to (\ref{Hform}), the CSI values for nearby sub-channels and antennas are correlated due to similar propagation paths, gains, delays and AoDs. This correlation will be exploited to reduce the CSI feedback. We note that this is a lossy compression problem, where the source samples follow a complicated correlation structure governed by (\ref{Hform}). Designing good practical codes for lossy compression is challenging even for memoryless sources with simpler underlying source distributions. Here, we will use a deep NN architecture, called DeepCMC, which uses CNN layers and entropy coding blocks to learn the CSI compression scheme that can best leverage the underlying correlations.

\section{DeepCMC}\label{section:2}

The overview of our proposed model architecture for encoding and subsequent reconstruction of the CSI feedback from user is shown in Fig.~\ref{single}, where the two channel inputs represent the real and imaginary parts of channel matrix. The user compresses its CSI into a variable length bit stream using the local encoder. The encoder at the user comprises a CNN-based feature encoder, a uniform element-wise scalar quantizer, and an entropy encoder. The feature encoder extracts key features from the CSI matrix to obtain a lower dimensional representation, which is subsequently converted into a discrete-valued vector by applying scalar quantization. While previous works simply send the 32-bit scalar quantized version of the feature vector as CSI feedback, we have observed that the CNN-based autoencoder structure does not result in a sequence of independent and uniformly distributed bits; and hence, can be further compressed. 

To further reduce the required feedback amount, we employ an entropy encoder; in particular, we use the context-adaptive binary arithmetic coding (CABAC) technique  \cite{marpe2003context}, which outputs a variable-length bit stream. Upon receiving this CSI-bearing bit stream, the BS first processes it by an entropy decoder to reproduce the lower-dimensional representation of the CSI feedback. This representation is then input to the feature decoder NN to reconstruct the estimated channel gain matrix. We present each component of our proposed model in more detail below. 
\subsection{Feature encoder and decoder}

\begin{figure}[t!]
\centering
\includegraphics[scale=0.4]{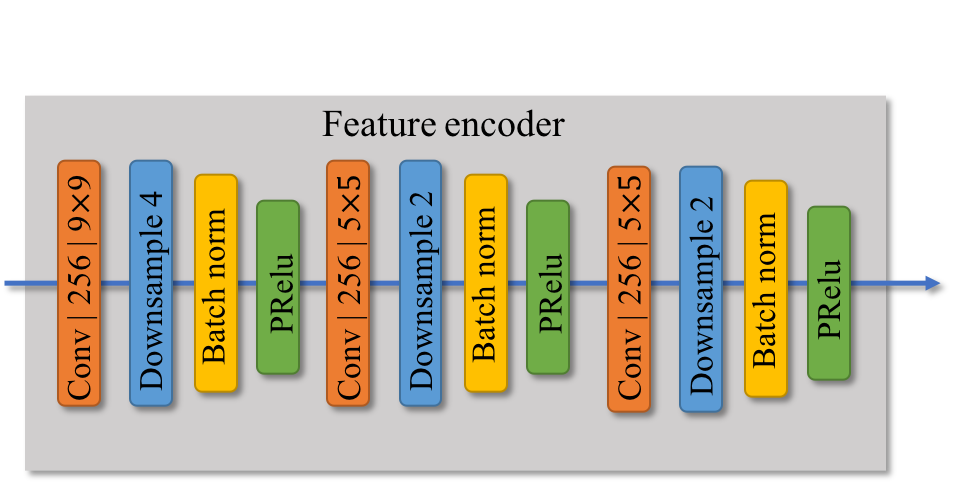}
\caption{Feature encoder architecture.}
\label{encoder}
\end{figure}

\begin{figure}[t!]
\centering
\includegraphics[scale=0.4]{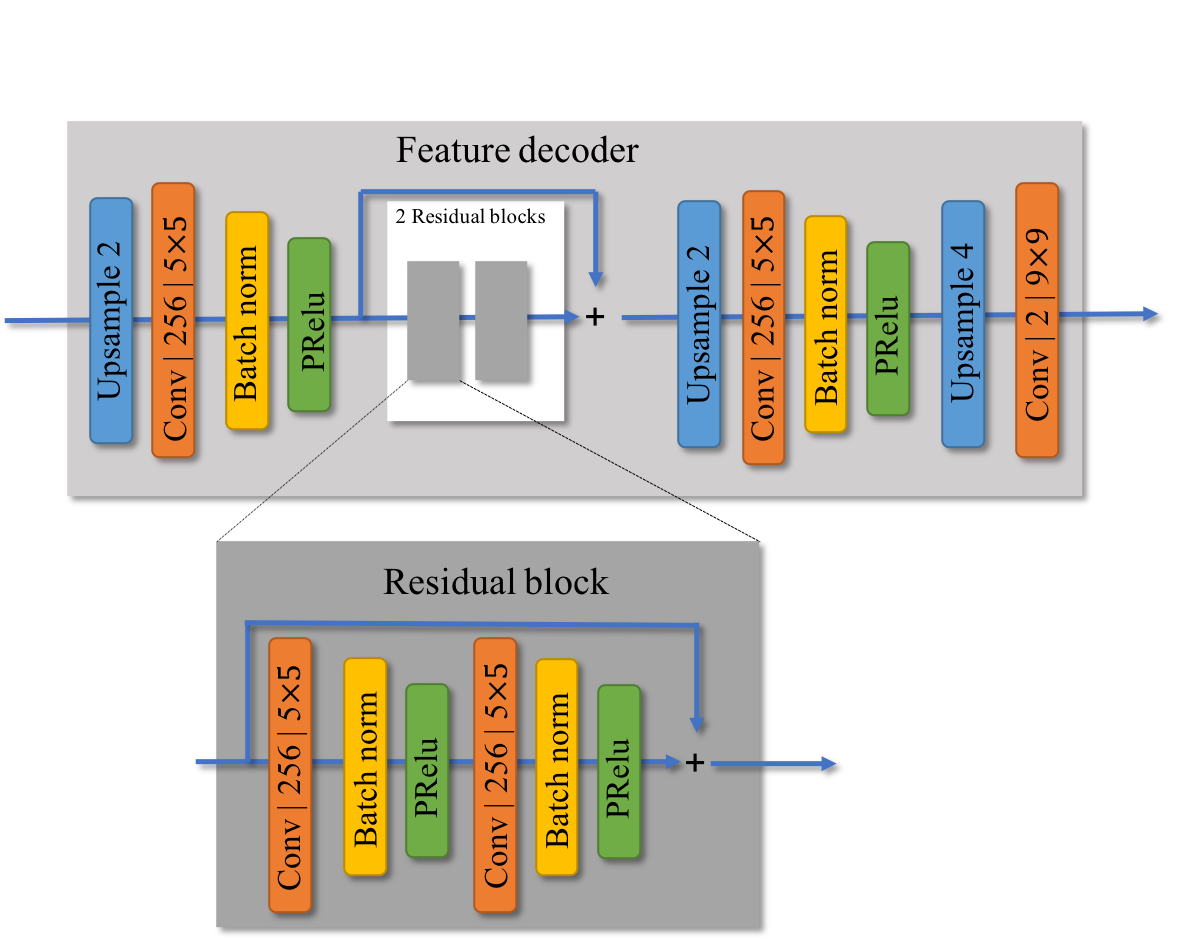}
\caption{Feature decoder architecture.}
\label{decoder}
\end{figure}

The CNN architecture used for the feature encoder and decoder are presented in Fig. \ref{encoder} and Fig. \ref{decoder}, respectively, where Conv$|256|9\times9$ represents a convolutional layer with 256 kernels, each of size $9\times9$. The feature encoder consists of three convolutional layers, the first of which uses kernels of size $9\times9$, and the other two use kernels of size $5\times5$. The ``SAME'' padding technique is used, such that the input and output of each convolutional layer have the same size (the number of channels vary). Each convolutional layer is followed by downsampling to reduce dimensionality. We use PRelu as the activation function, and apply batch normalization to each layer. Let
\small
\begin{equation}
\mathbf{M}=f_{\rm{f-en}}(\mathbf{H}, \Theta_{en}),   
\end{equation}
\normalsize
where $f_{\rm{f-en}}$ denotes the feature encoder at the user, and $\Theta_{en}$ denotes its parameter vector. $\mathbf{M}$ consists of $256$ feature maps of size $\frac{N_t}{16} \times \frac{N_c}{16}$. Note that this fully convolutional architecture allows us to use the same encoder network for any number of transmit antennas and subcarriers, while the feature vector dimension depends on the input size, which allows us to scale the CSI feedback volume with the channel dimension.

The feature decoder at BS performs the counterpart inverse operations, consisting of convolutional and upsampling layers. At the BS, the output of the entropy decoder is fed into the feature decoder to reconstruct the channel gain matrix. Similarly to the feature encoder, the decoder includes three layers of convolutions (with the same kernel sizes as the encoder) and upsampling (inverse of the downsampling operation at the encoder). The decoder architecture also includes two residual blocks with shortcut connections that skip several layers with $+$ denoting element-wise addition in Fig. \ref{decoder}. This structure eases the training of the network by preventing vanishing gradient along the stacked non-linear layers~\cite{he2016deep}. To enable this, the input and output of a residual block must have the same size. Each residual block comprises two convolutional layers (normalized using the batch norm) and uses PRelu as the activation function. Inspired by \cite{mentzer2018conditional}, we also use an identical shortcut connecting the input and output of the residual blocks, which improves the performance as revealed by the experiments. Let 
\small
\begin{equation}
\widehat{\mathbf{H}}=f_{\rm{f-de}}(\widehat{\mathbf{M}
}, \Theta_{de}),   
\end{equation}
\normalsize
be the output of the joint decoder where $\Theta_{de}$ is its set of parameters and $\widehat{\mathbf{M}}$ is the estimate of $\mathbf{M}$ provided by the entropy decoder. $\widehat{\mathbf{H}}$ denotes the reconstructed CSI feedback. 

\subsection{Quantization and Entropy coding}
A major contribution of our proposed model in comparison with the existing DNN architectures for CSI compression in the literature, such as CSINet\cite{wen2018deep}, is the inclusion of the entropy coding block which encodes quantized CSI data into bits at rates closely approaching the entropy. 

Quantization is performed by a uniform scalar quantizer denoted by $f_{q}$, which quantizes each element of $\mathbf{M}$ to the closest integer. We denote the quantized output as $\overline{\mathbf{M}}$, i.e.,
\small
\begin{equation}
\overline{\mathbf{M}}=f_{q}(\mathbf{M}).
\end{equation}
\normalsize

The entropy encoder converts the quantized values in $\overline{\mathbf{M}}$ into bit streams using CABAC \cite{marpe2003context} based on the input probability model learned during training. Let 
\small
\begin{equation}
\mathbf{s} = f_{\rm{e-en}}(\overline{\mathbf{M}}, P) \end{equation}
\normalsize
denote the bit stream derived by passing $\overline{\mathbf{M}}$ through the entropy coder, denoted by $f_{\rm{e-en}}$, where $P$ is the probability density function, estimated during training, as it will be described later in the following subsection.   

The estimate of $\mathbf{M}$, denoted by $\widehat{\mathbf{M}}$, is recovered at the BS by decoding the received codeword $\mathbf{s}$ using the corresponding entropy decoder as 
\small
\begin{equation}
\widehat{\mathbf{M}}= f_{\rm{e-de}}(\mathbf{s}, P).    
\end{equation}
\normalsize
Finally, $\widehat{\mathbf{M}}$ is fed into the feature decoder to reconstruct the CSI feedback.

\subsection{Optimization}
As quantization is not a differentiable function, it cannot be implemented within the gradient-based optimization framework. To overcome this, we replace the uniform scalar quantizer with independently and identically distributed (i.i.d) uniform noise during training. Hence, denoting the quantization noise vector by $\Delta\mathbf{M}$ with i.i.d elements from $U[0,1]$, we approximate the quantized feature matrix by 
\small
\begin{equation}
\widetilde{\mathbf{M}}=\mathbf{M}+\Delta\mathbf{M}. 
\end{equation}
\normalsize

Now denote by $P(\widetilde{\mathbf{M}}, \Theta_{p})$, the probability density function for $\widetilde{\mathbf{M}}$ specified by the set of parameters $\Theta_{p}$, which is estimated during training utilizing a similar technique as in \cite{balle2018variational}. Our loss function is given by
\small
\begin{align}\label{Loss}
&L(\Theta_{en}, \Theta_{de}, \Theta_{p}) = \nonumber \\  
& \mathbb{E}_{\mathbf{H}, \Delta \mathbf{M}}\Bigg(-\frac{1}{N_cN_t}\log P(f_{\rm{f-en}}(\mathbf{H}, \Theta_{en})+\Delta \mathbf{M}, \Theta_{p}) \nonumber \\ 
& + \lambda \text{MSE} \bigg(f_{\rm{f-de}}\big(f_{\rm{f-en}}(\mathbf{H}, \Theta_{en})+\Delta \mathbf{M}, \Theta_{de}\big), \mathbf{H}\bigg) \Bigg),   
\end{align}
\normalsize
where  
\small
\begin{equation}
\text{MSE} \bigg(\widehat{\mathbf{H}}, \mathbf{H}, \lambda\bigg)=\frac{1}{N_cN_t}(\widehat{\mathbf{H}}-\mathbf{H})^2,
\end{equation}
\normalsize
and the expectation is over the training set of channel matrices and the quantization noise. During training, the entropy of the quantized encoder outputs, estimated by the trainable probability model, is jointly minimized with the reconstruction MSE by optimizing the parameters for both the probability model and the autoencoder. By utilizing the entropy coding block with the optimized probability model, the actual bit rate of the encoder output closely approximates this entropy. More precisely, the first part of the loss function (\ref{Loss}) represents the entropy of the feedback data, or equivalently the size of feedback that must be transmitted, while the second part is the weighted mean square error (MSE) of the reconstructed channel gain matrices. Hence, training $\Theta_{en}, \Theta_{de}$ and $\Theta_{p}$ values, which parameterize the feature encoder, the feature decoder, and the probability models, respectively, minimizes the feedback overhead and the reconstruction loss, simultaneously. The $\lambda$ value governs the trade-off between the compression rate and the reconstruction loss. A larger $\lambda$ leads to a better reconstruction but a higher feedback overhead and vice versa. 

\begin{table}[!t] 
\centering
\caption{Performance of DeepCMC and CSINet schemes in terms of NMSE and cosine correlation for similar compression rate values (bits per channel dimension) ($N_c=256, N_t=32$).} \label{Single}
\resizebox{8.5cm}{!}{%
\begin{tabular}{c|c|c|c|c|c}
\footnotesize
Methods & $\lambda$ & Bit rate & Entropy & $\mathrm{NMSE}$ (dB)& $\rho$\\
\hline 
\multirow{5}*{DeepCMC} &$10^4$ &0.006068 & 0.003853 & -4.12 & 0.8401\\
\cline{2-6}
&$5 \times 10^4$ & 0.01353 & 0.01152 & -7.31 & 0.9337\\
\cline{2-6}
&$10^5$ & 0.01931 & 0.01808 & -9.20 & 0.9579\\
\cline{2-6}
& $5 \times 10^5$ & 0.05353 & 0.05478 & -11.83 &  0.9732\\
\cline{2-6}
& $10^6$ & 0.07658 & 0.07488 &  -12.45 & 0.9770\\
\cline{2-6}
& $5 \times 10^6$ & 0.1526 & 0.1509 & -13.57 & 0.9808\\
\hline 
\multirow{4}*{CSINet} & NA & 0.015625 & NA &  -1.31 & 0.6903\\
\cline{2-6}
& NA & 0.03125 & NA & -2.90 & 0.7806\\
\cline{2-6}
& NA & 0.0625 & NA & -5.33 & 0.8856\\
\cline{2-6}
& NA & 0.125 & NA &  -5.25 & 0.8783\\
\hline
\end{tabular}}
\end{table}
\normalsize
In order to recover the trade-off between the compression rate and the reconstruction loss, we train DeepCMC with different $\lambda$ values. For a small $\lambda$ value, the network tries to reduce the feedback rate, while as $\lambda$ increases, it tries to keep the MSE under control while slightly increasing the rate. After training, each $\lambda$ value specifies a set of parameters $\Theta_{en}, \Theta_{de}, \Theta_{p}$. By selecting the $\lambda$ value according to user's requirements in terms of CSI quality and the available feedback capacity, we can obtain the encoder and decoder parameters with the best performance under these constraints. This would require the user and the BS to have a list of encoder/decoder parameters to be used for different rate-MSE quality trade-offs, and the user to send the $\lambda$ value together with the encoded bitstream $\mathbf{s}$ to the BS, so that the BS employs the matching decoder parameters.

We emphasize here that the feature encoder and decoder networks are fully convolutional, and do not include any fully connected layers. Moreover the implemented entropy code can operate on inputs of any size. Therefore, the DeepCMC architecture can be trained on, or used for any channel matrix whose height and width are multiples of 16, since the feature encoder has a total downsampling rate of 16 (or, of any size, which can be made a multiple of 16 by padding). This is different from the existing NN-based CSI comrpession techqniues which are trained for a particular input size.

\section{Simulation Results}\label{sec4}

We use the COST 2100 channel model \cite{COST2100} to generate sample channel matrices for training and testing DeepCMC. We consider the indoor picocellular scenario at 5.3 GHz, where the BS is equipped with a ULA of dipole antennas positioned at the center of a $20m \times 20m$ square. The user is placed within this square uniformly at random. All other parameters follow the default settings in \cite{COST2100}. 


We use the normalized MSE (NMSE) and cosine correlation as the performance measures to compare our results with CSINet \cite{wen2018deep}. These measures are defined as follows,
\footnotesize
\begin{equation}
\mathrm{NMSE} \triangleq \mathbb{E}\left\{ \frac{ \|\mathbf{H}-\mathbf{\hat{H}}\|_2^2}{ \|\mathbf{H}\|_2^2} \right\},   
\end{equation}
\normalsize
and
\footnotesize
\begin{equation}
 \rho \triangleq \mathbb{E} \left\{\frac{1}{N_c} \sum^{N_c}_{n=1} \frac{|\mathbf{\hat{h}}_n^H \mathbf{h}_n|}{\|\mathbf{\hat{h}}_n\| \|\mathbf{h}_n\| }\right\}.   
\end{equation}
\normalsize

We first compare the performance of our DeepCMC scheme with CSINet. We train both models on the same data set of 80000 CSI realizations with $N_c=256$ and $N_t=32$. Table \ref{Single} provides the corresponding results tested on 20000 CSI realizations with same size as the training data. We train the DeepCMC architecture for different $\lambda$ values, which governs the trade-off between the compression rate and the reconstruction loss. We evaluate both the average entropy of the quantized outputs of the feature encoder with the test CSI matrices as input, i.e., $\overline{\mathbf{M}}$, and the average number of actual bits sent by the user. The latter includes the length of the bit streams generated by the entropy encoder with $\overline{\mathbf{M}}$ as the input and the value of $\lambda$ as a  $16$-bit integer, assuming that the user can decide on a different $\lambda$ value at each training instance. Naturally, the bit rate will further reduce if the BS and the user agree on a fixed $\lambda$ value throughout their operation. These two metrics are both normalized by $N_cN_t$, the total channel dimension. It is shown that the actual bit rate closely approximates the entropy of the quantized feature encoder outputs. The encoder output of CSINet is the feature vector, consisting of 32-bit float values. The length of this vector, denoted by $m$, determines the compression ratio, and hence the bit rate, of CSINet. We train CSINet for $m=8, 16, 32, 64$, which correspond to bit rates of $0.01562$, $0.03125$, $0.0625$, and $0.125$, respectively.

\begin{figure}[!t]
\centering
\includegraphics[width=0.82\linewidth]{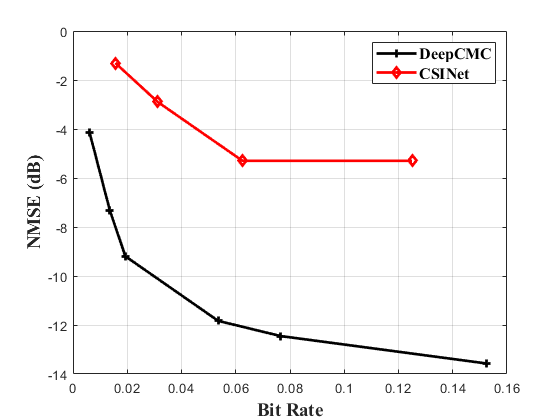}
\caption{Bit rate-NMSE trade-off of DeepCMC vs. CSINet, $N_c=256$, $N_t=32$.}\label{csinet}
\end{figure}

The bit rate-NMSE trade-offs achieved by DeepCMC and CSINet are plotted in Fig. \ref{csinet}. As it can be observed from Table \ref{single} and Fig. \ref{csinet}, DeepCMC provides significant improvement in the quality of the reconstructed CSI at the BS with respect to CSINet at all bit rate values. We remark here that, as reported in \cite{CSCSI1}, CSINet itself provides $3-6$~dB improvement in $\mathrm{NMSE}$ compared to other known CSI compression techniques in the literature exploiting the sparsity of the channel gain matrix. However, the gains from DeepCMC are even more drastic, achieving remarkably good reconstruction of the channel gain matrix with $\mathrm{NMSE}$ of $-13$~dB and $\rho$ equals to $0.98$ at a bit rate lower than $0.16$ bits per channel dimension. These results show that DeepCMC outperforms CSINet $4$ to $6$ dB in NMSE for the range of compression rates considered here. For example, for a target value of $\mathrm{NMSE}= -5$~dB, DeepCMC can provide more than 5 times reduction in the number of bits that must be fed back from the user to the BS. 

We remark here that, due to the critical nature of channel estimation at the BS, these feedback bits are typically transmitted using a low coding rate and a limited constellation size (e.g., BPSK or QPSK) to guarantee their correct decoding at the BS. This highlights the significance of reducing the required feedback bits, which can be directly translated into channel resource blocks \cite{LTE_feedback}. We further observe from Fig. \ref{csinet} that the NMSE of DeepCMC drops quite rapidly with bit rate, while CSINet has a much less significant reduction. This implies that DeepCMC better exploits the limited number of bits to capture the most essential information in the CSI data. 

These improvements are mainly due to the incorporation of the quantization layer into the training procedure, the introduction of the entropy coder, and the improved feature extraction architecture. The entropy coder with the approximate probability model greatly reduces the codeword lengths. Our experiments also reveal that the introduction of the shortcut connection across two residual blocks also improves the performance of DeepCMC, as well as using PRelu as the activation function.

\begin{table}[!t] 
\centering 
\caption{Performance of DeepCMC (trained with $\lambda=10^5$) for different number of subcarriers in the test channel with $N_t=32$}\label{Sizetable}
\footnotesize
\begin{tabular}{c|c|c|c|c}
$N_c$ & Bit rate & Entropy & $\mathrm{NMSE}$ (dB) & $\rho$\\
\hline 
 128 & 0.02192 & 0.02017 & -8.53 & 0.9542\\
\cline{1-5}
 160 & 0.02086 & 0.01933 & -8.76 & 0.9557\\
\cline{1-5}
 192 & 0.01877 & 0.02016 & -8.91 &  0.9569\\
\hline
 224 &
 0.01836 & 0.01966 & -9.03 & 0.9577\\
\cline{1-5}
 256 & 0.01808 & 0.01931 & -9.20 & 0.9579\\
\hline
\end{tabular}
\end{table}

\begin{figure}[!t]
\centering
\includegraphics[width=0.85\linewidth]{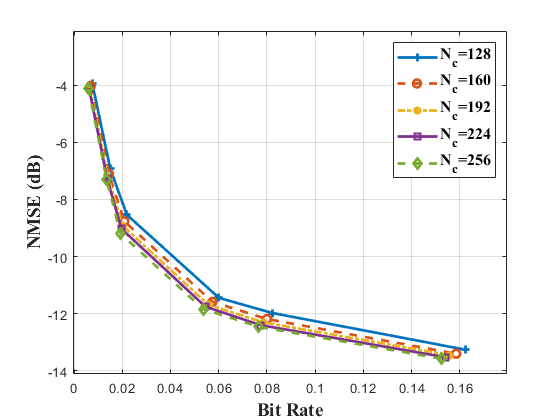}
\caption{Bit rate-NMSE trade-off for different number of subcarriers in the test channel with $N_t=32$.}\label{sizefigure}
\end{figure}

\normalsize
We then test DeepCMC (trained with $N_c=256$ and $N_t=32$) with different number of subcarriers, $N_c$. It is important that the feedback scheme is flexible in terms of the channel dimension, as the number of subcarriers or transmit antennas available may change from system to system, and even over time, e.g., different resources may be allocated to different sectors in a cell at different times. In our simulations, we keep the number of transmit antennas as $N_t=32$, and consider different number of subcarriers, $N_c=128, 160, 192, 224, 256$, each with $20000$ test samples. We summarized the performance of DeepCMC, trained with $\lambda=10^5$, in Table \ref{Sizetable}. We also present the bit rate-NMSE trade-off in Fig. \ref{sizefigure}, which is obtained by testing the DeepCMC network trained for different values of $\lambda$. Although the performance degrades as the difference between the test channel dimension and the training dimension increases; the degradation still remains negligible even when the test CSI realizations are only half of the size of the channel dimension DeepCMC has been trained on. As we see in Fig. \ref{sizefigure}, this holds for the whole range of bit rates, the loss due to training and test channel dimension mismatch being slightly larger at higher bit rates.


\section{Conclusions}\label{sec5}
In this paper, we proposed a convolutional deep learning architecture, called DeepCMC, for efficient compression of the channel gain matrix to reduce the significant CSI feedback load in massive MIMO systems. DeepCMC is composed of fully convolutional layers followed by quantization and entropy coding blocks, and outperforms state of the art DL-based CSI compression techniques, providing drastic improvements in CSI estimation quality at even extremely low feedback rates. Another advantage of the DeepCMC architecture is its flexibility in terms of the channel dimension; that is, the same architecture can be used for CSI feedback over channels with different numbers of transmit antennas or subcarriers, significantly increasing its practical applicability.

\bibliographystyle{IEEEtran}
\bibliography{main}

\begin{thebibliography}{10}
\providecommand{\url}[1]{#1}
\csname url@samestyle\endcsname
\providecommand{\newblock}{\relax}
\providecommand{\bibinfo}[2]{#2}
\providecommand{\BIBentrySTDinterwordspacing}{\spaceskip=0pt\relax}
\providecommand{\BIBentryALTinterwordstretchfactor}{4}
\providecommand{\BIBentryALTinterwordspacing}{\spaceskip=\fontdimen2\font plus
\BIBentryALTinterwordstretchfactor\fontdimen3\font minus
  \fontdimen4\font\relax}
\providecommand{\BIBforeignlanguage}[2]{{%
\expandafter\ifx\csname l@#1\endcsname\relax
\typeout{** WARNING: IEEEtran.bst: No hyphenation pattern has been}%
\typeout{** loaded for the language `#1'. Using the pattern for}%
\typeout{** the default language instead.}%
\else
\language=\csname l@#1\endcsname
\fi
#2}}
\providecommand{\BIBdecl}{\relax}
\BIBdecl

\bibitem{CodebookCSI1}
D.~J. {Love}, R.~W. {Heath}, V.~K. {N. Lau}, D.~{Gesbert}, B.~D. {Rao}, and
  M.~{Andrews}, ``An overview of limited feedback in wireless communication
  systems,'' \emph{{IEEE} J. Sel. Areas Commun.}, vol.~26, no.~8, pp.
  1341--1365, Oct. 2008.

\bibitem{CodebookCSI2}
H.~{Shirani-Mehr} and G.~{Caire}, ``Channel state feedback schemes for
  multiuser {MIMO-OFDM} downlink,'' \emph{{IEEE} Trans. Commun.}, vol.~57,
  no.~9, pp. 2713--2723, Sep. 2009.

\bibitem{CSCSI1}
P.~{Kuo}, H.~T. {Kung}, and P.~{Ting}, ``Compressive sensing based channel
  feedback protocols for spatially-correlated massive antenna arrays,'' in
  \emph{IEEE Wireless Commun. and Netw. Conf. (WCNC)}, April 2012, pp.
  492--497.

\bibitem{CSCSI2}
X.~{Rao} and V.~K.~N. {Lau}, ``Distributed compressive {CSIT} estimation and
  feedback for {FDD} multi-user massive {MIMO} systems,'' \emph{IEEE
  Transactions on Signal Processing}, vol.~62, no.~12, pp. 3261--3271, June
  2014.

\bibitem{MLintheAir}
\BIBentryALTinterwordspacing
D.~G{\"{u}}nd{\"{u}}z, P.~de~Kerret, N.~D. Sidiropoulos, D.~Gesbert, C.~Murthy,
  and M.~van~der Schaar, ``Machine learning in the air,'' \emph{CoRR}, vol.
  abs/1904.12385, 2019. [Online]. Available:
  \url{http://arxiv.org/abs/1904.12385}
\BIBentrySTDinterwordspacing

\bibitem{wen2018deep}
C.-K. Wen, W.-T. Shih, and S.~Jin, ``Deep learning for massive {MIMO} {CSI}
  feedback,'' \emph{{IEEE} Wireless Commun. Lett.}, vol.~7, no.~5, pp.
  748--751, 2018.

\bibitem{DLCSI3}
C.~{Lu}, W.~{Xu}, H.~{Shen}, J.~{Zhu}, and K.~{Wang}, ``{MIMO} channel
  information feedback using deep recurrent network,'' \emph{IEEE Commun.
  Lett.}, vol.~23, no.~1, pp. 188--191, Jan 2019.

\bibitem{DLCSI4}
Z.~{Liu}, L.~{Zhang}, and Z.~{Ding}, ``Exploiting bi-directional channel
  reciprocity in deep learning for low rate massive mimo csi feedback,''
  \emph{IEEE Wireless Commun. Lett.}, pp. 1--1, 2019.

\bibitem{HMIMO}
Z.~{Gao}, C.~{Hu}, L.~{Dai}, and Z.~{Wang}, ``Channel estimation for
  millimeter-wave massive {MIMO} with hybrid precoding over frequency-selective
  fading channels,'' \emph{IEEE Communications Letters}, vol.~20, no.~6, pp.
  1259--1262, June 2016.

\bibitem{marpe2003context}
D.~Marpe, H.~Schwarz, and T.~Wiegand, ``Context-based adaptive binary
  arithmetic coding in the h. 264/avc video compression standard,'' \emph{IEEE
  Trans. Circuits and Syst. for Video Tech.}, vol.~13, no.~7, pp. 620--636,
  2003.

\bibitem{he2016deep}
K.~He, X.~Zhang, S.~Ren, and J.~Sun, ``Deep residual learning for image
  recognition,'' in \emph{Proc. {IEEE} Int'l Conf. Comp. vision and pattern
  recognition (CVPR)}, Las Vegas, NV, Jun. 2016, pp. 770--778.

\bibitem{mentzer2018conditional}
F.~Mentzer, E.~Agustsson, M.~Tschannen, R.~Timofte, and L.~Van~Gool,
  ``Conditional probability models for deep image compression,'' in \emph{Proc.
  {IEEE} Int'l Conf. Comp. vision and pattern recognition (CVPR)}, Salt Lake
  City, UT, Jun 2018, pp. 4394--4402.

\bibitem{balle2018variational}
J.~Ball{\'e}, D.~Minnen, S.~Singh, S.~J. Hwang, and N.~Johnston, ``Variational
  image compression with a scale hyperprior,'' \emph{arXiv: 1802.01436v2
  [eess.IV]}, May 2018.

\bibitem{COST2100}
L.~{Liu}, C.~{Oestges}, J.~{Poutanen}, K.~{Haneda}, P.~{Vainikainen},
  F.~{Quitin}, F.~{Tufvesson}, and P.~D. {Doncker}, ``The {COST} 2100 {MIMO}
  channel model,'' \emph{IEEE Wireless Commun.}, vol.~19, no.~6, pp. 92--99,
  December 2012.

\bibitem{LTE_feedback}
A.~{Chiumento}, C.~{Desset}, S.~{Pollin}, L.~{Van der Perre}, and
  R.~{Lauwereins}, ``Impact of {CSI} feedback strategies on lte downlink and
  reinforcement learning solutions for optimal allocation,'' \emph{IEEE Trans.
  Vehicular Tech.}, vol.~66, no.~1, pp. 550--562, Jan 2017.

\end{thebibliography}
\end{document}